\documentclass[11pt]{article}
\usepackage{latexsym}
\usepackage{amssymb}
\usepackage{epsf}
\usepackage{amsmath}
\usepackage{slashed}
\usepackage[hypertex]{hyperref}
\usepackage{graphicx}% Include figure files

\newcommand{\psl}{\mathbf{p} \hspace{-0.5 em}/}
\newcommand{\psll}{\mathbf{P} \hspace{-0.5 em}/}
\newcommand{\ptsl}{\mathrm{p} \hspace{-0.5 em}/_T}

\begin{document}
\pagestyle{empty}

%\begin{flushright}
%\today 
%\end{flushright}

\hfill TU-874

\vspace{3cm}

\begin{center}

{\bf\LARGE 
%$Z^{\prime}$ measurements in Higgsless models at the LHC
Measurements of neutral vector resonance in Higgsless models at the LHC}
\\

\vspace*{1.5cm}
{\large 
Masaki Asano$^a$ and Yasuhiro Shimizu$^{a,b}$
} \\
\vspace*{0.5cm}

$^a${\it Department of Physics, Tohoku University, Sendai 980-8578,
 Japan}\\
$^b${\it IIAIR, Tohoku University, Sendai 980-8578,
 Japan}\\
\vspace*{0.5cm}

\end{center}

\vspace*{1.0cm}

\begin{abstract}
{\normalsize
In Higgsless models, new vector resonances appear to 
restore the unitarity of the $W_L W_L$ scattering amplitude 
without the Higgs boson.
In the ideal delocalized three site Higgsless model,
one of large production cross section of the neutral vector 
resonance ($Z^{\prime}$)  at the Large Hadron Collider is 
the $W$-associated production, $pp \to Z^{\prime}W^{\pm} \to W^{\mp}W^{\pm}W^{\pm}$.
Although the dileptonic decay channel, 
$l \nu l^{\prime} \nu^{\prime} jj$, is experimentally clean 
to search for the $Z'$ signals, it is difficult to reconstruct 
the $Z^{\prime}$ invariant mass due to the two neutrinos in the 
final state.
We study collider signatures of $Z^{\prime}$ using the
$M_{T2}$-Assisted On-Shell (MAOS) reconstruction of the missing neutrino 
momenta. We show the prospect of the $Z^{\prime}$ mass determination
in the channel, $l \nu l^{\prime} \nu^{\prime} jj$, at the Large 
Hadron Collider.

}
\end{abstract} 

%%%%%%%%%%%%%%%%%%%%%%%%%%%%%%%%%%%%%%%%%%%%%%%%%%%%%%%%%%%%%%%%%%%%%%%%%%%%
\newpage
\baselineskip=18pt
\setcounter{page}{2}
\pagestyle{plain}
\baselineskip=18pt
\pagestyle{plain}

\setcounter{footnote}{0}

%%%%%%%%%%%%%%%%%%%%%%%%%%%%%%%%%%%%%%%%%%%%%%%%%%%%%%%%%%%%%%%%%%%%%%%%%%%%%%%%%
\section{ Introduction }\label{sect:intro}
%%%%%%%%%%%%%%%%%%%%%%%%%%%%%%%%%%%%%%%%%%%%%%%%%%%%%%%%%%%%%%%%%%%%%%%%%%%%%%%%%

The Large Hadron Collider (LHC) experiments are now operating. At the LHC, 
a new electroweak signal will be discovered because the unitarity of the 
$W_L W_L$ scattering amplitude is violated at the scale higher around 1 TeV 
due to the $E^2$ dependence. 
%One of the candidate is the Higgs boson which is a scalar 
%$SU(2)_L$ doublet. 
One of the candidates is a signal from a scalar $SU(2)_L$ doublet, 
called the Higgs boson. Due to the contribution from the 
Higgs boson, the $E^2$ dependence of the $W_L W_L$ scattering amplitude 
vanishes in the standard model (SM). 

On the other hand, it is also possible to maintain 
the unitarity without the Higgs scalar in Higgsless models
%there are other plausible possibilities, called Higgsless models 
\cite{Csaki:2003dt, Csaki:2003zu}.
In Higgsless models, vector resonances which are responsible for 
the restoration of the unitarity of the $W_L W_L$ scattering amplitude 
will be produced at the LHC.
The masses are less than around 1 TeV to avoid the unitarity violation, but
such particles are strictly constrained by direct observation if their 
couplings to SM fermions are similar to the SM ones \cite{Amsler:2008zzb}.
Also the coupling induces a large correction to electroweak precision 
measurements at the tree level. 
To avoid the constraints, the couplings should be suppressed and
%In extra dimensional Higgsless scenario, it can be explained by the fermion 
%delocalization[]. 
it is realized by the fermion delocalization \cite{Cacciapaglia:2004rb, Foadi:2004ps, 
Chivukula:2005bn, Casalbuoni:2005rs, Chivukula:2005xm, Chivukula:2005cc}. 
%It is also possible to explain results of electroweak precision measurements.
The vector resonances therefore have only small coupling to the SM fermions in 
Higgsless models.

%The mass determination of the vector resonances at the LHC in Higgsless
%models has already been studied by many authors 
Fermiophobic vector resonances can be produced mainly by the weak gauge boson 
associated process or the vector boson fusion process and mainly decay into 
the weak gauge bosons at the LHC. The measurements of charged vector resonance  
have been studied \cite{Birkedal:2004au,He:2007ge,Alves:2008up,Bian:2009kf,Han:2009qr}. 
However, in a case that the $W$-associated production is dominant, the mass 
determination of the neutral vector resonance ($Z^{\prime}$) is very difficult 
because there are three $W$ bosons in the final state. A way using 
$W^{\pm}W^{\pm}W^{\mp} \to l^{\pm} \nu l^{\prime \pm} \nu^{\prime} jj$ mode 
is proposed by Tao Han et.al.\cite{Han:2009qr}, and they showed that the mass 
can be read off by an endpoint of the $m(l j j)$ 
distribution from the parton level study.

In this paper, we show that neutral vector resonance can be
reconstructed even in 
$W W W  \to l  \nu l^{\prime} \nu^{\prime} jj$ 
final state using $M_{T2}$ Assisted On-Shell (MAOS) reconstruction of 
missing momenta \cite{Cho:2008tj, Choi:2009hn,Choi:2010dw}.
Since MAOS momenta equal true neutrino four momenta at the $M_{T2}$ 
\cite{Lester:1999tx} 
endpoint of $W W  \to l  \nu l^{\prime} \nu^{\prime}$ 
system, the $Z^{\prime}$ mass will be determined by the peak of the invariant 
mass of reconstructed $W^{\pm} W^{\mp}$ bosons.

This article is organized as follows. In the next section, we briefly review 
the Higgsless models. 
%%%%%%%%%%%%%%%%%%%
In our study, we focus on the three site Higgsless model
\cite{Chivukula:2006cg} as a benchmark model.
%%%%%%%%%%%%%%%%%%%
%we search the origin of masses of the standard model (SM) particles. 
In section 3, we briefly review the MAOS momenta. The Monte Carlo simulation 
study for measurements of neutral vector resonance at the LHC are discussed 
in section 4, where we show the mass will be measured with an integrated 
luminosity of $100$ ${\rm fb}^{-1}$ at $\sqrt{s} = 14 {\rm TeV}$.
%${\mathcal{L} = 100 {\rm fb}^{-1}}$ 
Section 5 is devoted to summary and discussion.

%%%%%%%%%%%%%%%%%%%%%%%%%%%%%%%%%%%%%%%%%%%%%%%%%%%%%%%%%%%%%%%%%%%%%%%%%%%%%%%%%
\section{ Higgsless models }\label{sect:model}
%%%%%%%%%%%%%%%%%%%%%%%%%%%%%%%%%%%%%%%%%%%%%%%%%%%%%%%%%%%%%%%%%%%%%%%%%%%%%%%%%

In continuum five-dimensional gauge theory, the cancellations of $E^2$ and $E^4$
dependence of $W_L W_L$ scattering amplitude are guaranteed by following sum 
rules \cite{Csaki:2003dt, SekharChivukula:2008mj}:
%--------------------------------------------------------->>>
\begin{eqnarray}
    \sum^{\infty}_{i=1} g_{Z_i WW}^2
&=& g_{WWWW} - g_{ZWW}^2 - g_{\gamma WW}^2, 
\\ \nonumber
  3 \sum^{\infty}_{i=1} g_{Z_i WW}^2 M_{Z_i}^2
&=& 4 g_{WWWW} M_W^2 - 3 g_{ZWW}^2 M_Z^2,
    \label{fig:sum_rule_1}
\end{eqnarray}
%---------------------------------------------------------<<<
where $g_{WWWW}$, $g_{ZWW}$ and $g_{\gamma WW}$ are the SM $WWWW$, $WWZ$ and 
$\gamma WW$ coupling constants, respectively. The $Z_i$ represents 
the $i$-th KK excitation and $Z_1$ is also denoted $Z^{\prime}$ in this paper. 
The $Z^{\prime}$ mass should be less than the unitarity violation scale 
$\sqrt{8 \pi} v \sim $ 1.2 TeV if there is no Higgs boson nor 
other vector resonances.

In this paper, we study the collider signatures of 
the three site Higgsless model \cite{Chivukula:2006cg} as a benchmark model
of Higgsless models. The model contains many essential ingredients of 
Higgsless models and a gauge invariant  four-dimensional effective theory 
of Higgsless models. 
We briefly review the three site Higgsless model.

%%%%%%%%%%%%%%%%%%%%%%%%%%%%%%%%%%%%%%%%%%%%%%%%%%%%%%%%%%%%%%%%%%%%%%%%%%%%%%%%%
%\subsection{Three site Higgsless model}
%%%%%%%%%%%%%%%%%%%%%%%%%%%%%%%%%%%%%%%%%%%%%%%%%%%%%%%%%%%%%%%%%%%%%%%%%%%%%%%%%

The three site Higgsless model is a deconstructed Higgsless model with 
only three sites
\footnote{ The gauge sector is equivalent to the Breaking Electroweak 
           Symmetry Strongly (BESS) model 
           \cite{Casalbuoni:1985kq, Casalbuoni:1995qt} one.}.
The model is based on two nonlinear $( SU(2) \times SU(2) )/SU(2)$ sigma models. 
The non-linear sigma fields, $U_1$ and $U_2$, are given as
%--------------------------------------------------------->>>
\begin{eqnarray}
 U_i &=& e^{i \pi_i^a \tau^a / f_i}  \quad {\rm for } \ i=1,2,
    \label{fig:U_i}
\end{eqnarray}
%---------------------------------------------------------<<<
where $\pi_i^a$ are Nambu-Goldstone bosons and $\tau^a$ are the Pauli matrices.
The model incorporates $SU(2) \times SU(2) \times U(1)$ gauge symmetry 
with gauge coupling strengths $g_0$, $g_1$ and $g_2$, respectively. The vacuum 
expectation values (VEVs) $f_1$ and $f_2$ break the 
$SU(2) \times SU(2) \times U(1)$ gauge symmetry to $U(1)_{em}$:
%--------------------------------------------------------->>>
\begin{eqnarray}
 \mathcal{L}_{gauge}
&=& \sum_{i=1}^2 \frac{f_i^2}{4} 
        {\rm Tr} \left[ (D_{\mu}U_i)^{\dagger} (D^{\mu}U_i) \right]
  - \sum_{i=0}^2 \frac{1}{2} 
        {\rm Tr} \left[ {\bf V}_{i \mu \nu} {\bf V}_i^{\mu \nu} \right],
\\ \nonumber
D_{\mu}U_i
&\equiv& \partial_{\mu} U_i 
         + i g_{(i-1)} {\bf V}_{(i-1) \mu} U_i - i g_i U_i {\bf V}_{i \mu}, 
\qquad 
{\bf V}_{i \mu} = \sum_{a=1,2,3} \frac{\tau^a}{2} V_{i \mu}^a 
\quad {\rm for } \ i=0,1,
\\ \nonumber
{\bf V}_{i \mu \nu}
&\equiv&  \partial_{\mu} {\bf V}_{i \nu}
        - \partial_{\nu} {\bf V}_{i \mu}
        + i g_i  \left[ {\bf V}_{i \mu}, {\bf V}_{i \nu}\right],
\qquad 
{\bf V}_{2 \mu} = \frac{\tau^3}{2} V_{2 \mu}^3,
    \label{fig:L_gauge}
\end{eqnarray}
%---------------------------------------------------------<<<
where the $V_{0(1)}^a$ is the gauge field for the $SU(2)$ gauge group 
at the site 0(1) and $V_2^3$ is the gauge field for the $U(1)$ 
gauge group at the site 2 which is embedded as the $\tau_3$ generator 
of $SU(2)$. 
Therefore, this model contains charged and neutral gauge bosons, 
$W^{\prime}$ and $Z^{\prime}$, in addition to the SM $W$, $Z$ and photon. 
The combinations of $V_{0,1}^{1,2}$ correspond to the $W^{\prime}$ 
and $W$. On the other hand, the combination of $V_{0,1,2}^3$ are the 
$Z^{\prime}$, $Z$ and photon, respectively. 
For simplicity, we take $f_1 = f_2 = \sqrt{2} v$ in this study. 

The left-handed fermions $\psi_{L0}$ and $\psi_{L1}$ are $SU(2)$ doublets 
coupling to the groups at the sites 0 and 1, respectively. 
On the other hand, the right handed fermions $\psi_{R1}$ are
SU(2) doublets coupling to the group at  
site 1 and $u_{R2}$, $d_{R2}$ and $e_{R2}$ are singlets coupling to the group 
at site 2.
The quantum numbers of the fermions are shown in Table \ref{tab:quantum_num}.
%===============================================================================
\begin{table*}[t]
\center{
  \begin{tabular}{|c||c|c|c|}
  \hline
  Particles        & $SU(2)_0$ & $SU(2)_1$ & $U(1)_2$ \\
  \hline
  $q_{L0}$, $l_{L0}$ & {\bf 2} & {\bf 1} & $1/6$, $-1/2$ \\
  \hline
  $q_{L1}$, $l_{L1}$ & {\bf 1} & {\bf 2} & $1/6$, $-1/2$ \\
  \hline
  $q_{R1}$, $l_{R1}$ & {\bf 1} & {\bf 2} & $1/6$, $-1/2$ \\
  \hline
  $u_{R2}$, $d_{R2}$, $e_{R2}$ & {\bf 1} & {\bf 1} & $2/3$, $-1/3$, $-1$ \\
  \hline
  \end{tabular}
}
  \caption{\small Quantum numbers of fermion fields.}
  \label{tab:quantum_num}
\end{table*}
%===============================================================================
The Yukawa couplings are written by 
%--------------------------------------------------------->>>
\begin{eqnarray}
 \mathcal{L}_{Yukawa}
&=& \lambda f_1 \bar{\psi}_{L0} U_1 \psi_{R1} 
   +        f_2 \bar{\psi}_{L1} U_2 
  \begin{pmatrix}
    \lambda^{\prime}_u & 0                  \\ 
     0                 & \lambda^{\prime}_d
  \end{pmatrix}
  \begin{pmatrix}
     u_{R2} \\ d_{R2}
  \end{pmatrix}
   + M \bar{\psi}_{L1}\psi_{R1}
   + {\rm h.c.}
\\ \nonumber
&=& 
  \begin{pmatrix}
     \bar{\psi}_{L0} & \bar{\psi}_{L1}
  \end{pmatrix}
       M
  \begin{pmatrix}
     \varepsilon_L & 0                  \\ 
      1            & \varepsilon_{uR,dR}
  \end{pmatrix}
  \begin{pmatrix}
     \psi_{R1} \\ u_{R2},d_{R2}
  \end{pmatrix}
  + {\rm h.c.} ,
    \label{fig:L_Yukawa}
\end{eqnarray}
%---------------------------------------------------------<<<
where $M \varepsilon_L = \lambda f_1 \equiv m$ and 
$M \varepsilon_{uR,dR} = \lambda^{\prime}_{u,d} f_2 \equiv m^{\prime}_{u,d}$.
In this paper, we assume that the $m$ and $M$ are universal for the generation 
and quark-lepton.

In the limit, $g_0/g_1 \ll 1$ and $g_2/g_1 \ll 1$, the gauge boson masses are 
given by
%--------------------------------------------------------->>>
\begin{eqnarray}
M_W
&\sim& \frac{g_0^2}{4} v^2 
       \left[ 1 - \frac{1}{4} \left( \frac{g_0}{g_1} \right)^2 \right],
\quad
M_{W^{\prime}} 
 \sim g_1^2 v^2 
       \left[ 1 + \frac{1}{4} \left( \frac{g_0}{g_1} \right)^2 \right],
\\ \nonumber
M_Z
&\sim& \frac{g_0^2}{4 c^2} v^2 
       \left[ 1 - \frac{(c^2 - s^2)^2}{4c^2} 
                  \left( \frac{g_0}{g_1} \right)^2 \right],
\quad
M_{Z^{\prime}} 
 \sim g_1^2 v^2 
       \left[ 1 + \frac{1}{4 c^2} \left( \frac{g_0}{g_1} \right)^2 \right].
    \label{fig:mass_gauge}
\end{eqnarray}
Here $s=\sin\theta$, $c=\cos\theta$, and $\tan\theta=g_2/g_0$.
%---------------------------------------------------------<<<
In the $\varepsilon_L \ll 1$ limit, fermion masses are written as the following:
%--------------------------------------------------------->>>
\begin{eqnarray}
m_{t,b}
&\sim& \frac{ m m_{t,b}^{\prime} }{ \sqrt{M^2 + m_{t,b}^{\prime 2}} },
\quad
M_{T,B} \sim \sqrt{M^2 + m_{t,b}^{\prime 2}},
    \label{fig:mass_fermion}
\end{eqnarray}
%---------------------------------------------------------<<<
where we show the masses of the third-generation quarks as an example. 
The $M_T(M_B)$  is the new heavy top(bottom) quark mass.

The couplings between the light fermions and the heavy gauge bosons are 
constrained by direct heavy gauge boson search and the electroweak 
precision measurements.
The coupling can be small by delocalizing fermion and $\varepsilon_L$ is a 
parameter which denotes the degree of the delocalization. In the case, 
$\varepsilon_L = (1 + \varepsilon_{fR}^2)^2
                    \{(g_0/g_1)^2/2 + \mathcal{O}((g_0/g_1)^4) + \cdots\}$, 
the coupling constants are given by 
%--------------------------------------------------------->>>
\begin{eqnarray}
g_{ffW^{\prime}}
&=& 0.
%\quad
%g_{ffZ^{\prime}} = ...
    \label{fig:couping_ffZ}
\end{eqnarray}
%---------------------------------------------------------<<<
This case is called ideal delocalization in which the precision electroweak 
corrections are minimized at tree level \cite{Chivukula:2005xm}
\footnote{ 
At the 1-loop level, the parameter does not satisfy the electroweak precision 
measurements \cite{Abe:2008hb}.
But we take the parameter in the following analysis since the 
correction to the signal in our collider study is small even if we take 
a parameter which satisfies the electroweak precision measurements 
at the 1-loop level.
}.
In the ideal delocalization case, the lower bound of $W^\prime$ mass, $380$ GeV, 
is given by the bound on the triple gauge vertex from the LEP-II experiments
\cite{Chivukula:2006cg}.

%%%%%%%%%%%%%%%%%%%%%%%%%%%%%%%%%%%%%%%%%%%%%%%%%%%%%%%%%%%%%%%%%%%%%%%%%%%%%%%%%
\section{ MAOS momentum  }\label{sect:MAOS}
%%%%%%%%%%%%%%%%%%%%%%%%%%%%%%%%%%%%%%%%%%%%%%%%%%%%%%%%%%%%%%%%%%%%%%%%%%%%%%%%%
In this section, we briefly review the MAOS momentum 
\cite{Cho:2008tj, Choi:2009hn,Choi:2010dw}. 
The MAOS momentum is defined using the $M_{T2}$ formula \cite{Lester:1999tx}. In the 
$WW \to l(p) \nu(k) l^{\prime}(p^{\prime}) \nu^{\prime}(k^{\prime}) $ 
system, the $M_{T2}$ valuable is written by 
%--------------------------------------------------------->>>
\begin{eqnarray}
M_{T2}^2
&=& \min_{ {\bf k}_T + {\bf k}_T^{\prime} = \psl_T } 
    \left[ \max \left\{ M_T^2 ({\bf p}_T,{\bf k}_T), 
                        M_T^2 ({\bf p}_T^{\prime},{\bf k}_T^{\prime})\right\} \right],
    \label{fig:MT2_1}
\end{eqnarray}
%---------------------------------------------------------<<<
where $\psl_T$ is the missing transverse momentum and the transverse mass, 
$M_T$, is defined by
%--------------------------------------------------------->>>
\begin{eqnarray}
 M_T^2 ({\bf p}_T,{\bf k}_T)
&=& 2 \left( |{\bf p}_T||{\bf k}_T| 
            - {\bf p}_T \cdot {\bf k}_T
     \right),
    \label{fig:MT2_2}
\end{eqnarray}
%---------------------------------------------------------<<<
where the lepton masses are neglected.

The MAOS momenta, $k^{MAOS}$ and $k^{\prime MAOS}$, are defined by the following.
Assuming $\psl_T = - ( {\bf p}_T + {\bf p}_T^{\prime} )$, the transverse momenta 
of $k^{MAOS}$ and $k^{\prime MAOS}$ are defined by
%--------------------------------------------------------->>>
\begin{eqnarray}
{\bf k}_T^{MAOS} &=& - {\bf p}_T^{\prime},
\quad
{\bf k}_T^{\prime MAOS}  =  - {\bf p}_T,
    \label{fig:MAOS_2}
\end{eqnarray}
%---------------------------------------------------------<<<
and
%--------------------------------------------------------->>>
\begin{eqnarray}
M_{T2}( {\bf p}_T,{\bf p}_T^{\prime},\psl_T) 
&=& \sqrt{2}\sqrt{ \left( |{\bf p}_T||{\bf k}_T^{MAOS}| 
                 - {\bf p}_T \cdot {\bf k}_T^{MAOS}
          \right)}, \\ \nonumber
&=& \sqrt{2}\sqrt{ \left( |{\bf p}_T^{\prime}||{\bf k}_T^{\prime MAOS}| 
                 - {\bf p}_T^{\prime} \cdot {\bf k}_T^{\prime MAOS}
          \right)}.
    \label{fig:MAOS_1}
\end{eqnarray}
%---------------------------------------------------------<<<
The MAOS momenta are also required to satisfy the following conditions in the 
$WW \to l(p) \nu(k) l^{\prime}(p^{\prime}) \nu^{\prime}(k^{\prime}) $ system:
%--------------------------------------------------------->>>
\begin{eqnarray}
    \left( k^{MAOS} \right)^2
&=& \left( k^{\prime MAOS} \right)^2 = 0, \\ \nonumber
    \left( p + k^{MAOS} \right)^2
&=& \left( p^{\prime} + k^{\prime MAOS} \right)^2 = M_{T2}^2.
    \label{fig:MAOS_3}
\end{eqnarray}
From these equations, the longitudinal momenta of $k^{MAOS}$ and 
$k^{\prime MAOS}$ are also determined by
%--------------------------------------------------------->>>
\begin{eqnarray}
    k_L^{MAOS} (\pm)
&=& \frac{|{\bf k}_T^{MAOS}|}{|{\bf p}_T|} p_L, 
\quad
    k_L^{\prime MAOS} (\pm)
 = \frac{|{\bf k}_T^{\prime MAOS}|}{|{\bf p}_T^{\prime}|} p_L^{\prime}.
    \label{fig:MAOS_4}
\end{eqnarray}
At the $M_{T2}$ endpoint, the MAOS momenta become equal to the final state 
neutrino momenta $k, k^{\prime}$, respectively.

In the next section, we show that the fermiophobic $Z^{\prime}$ can be 
reconstructed using MAOS momenta of neutrinos at the LHC through Monte Carlo 
simulations. The $Z^{\prime}$ mass can be measured by the invariant mass 
peak of the reconstructed $WW$.

%%%%%%%%%%%%%%%%%%%%%%%%%%%%%%%%%%%%%%%%%%%%%%%%%%%%%%%%%%%%%%%%%%%%%%%%%%%%%%%%%
\section{ Monte Carlo Simulation }\label{sect:simulation}
%%%%%%%%%%%%%%%%%%%%%%%%%%%%%%%%%%%%%%%%%%%%%%%%%%%%%%%%%%%%%%%%%%%%%%%%%%%%%%%%%
We study the possibility of $Z^{\prime}$ reconstruction through Monte Carlo 
simulations in this section. We investigate the ideal 
delocalized three site Higgsless model as a benchmark model and the parameters 
at the representative points are summarized in Table 
\ref{tab:representative points}.
%===============================================================================
\begin{table*}[t]
\center{
  \begin{tabular}{|c||c|c|c|c|c|c|}
  \hline
  $m_{Z^{\prime}}$ [GeV] &$m_{W^{\prime}}$ [GeV] 
% & $\epsilon_L$  & $\epsilon_R$ 
  & $M$ [GeV] 
  & $|g_{Z'WW}|$ 
  & $\sigma(Z^\prime W)$ [fb]
  & $\sigma(Z^\prime qq)$ [fb]
  & Br($Z^{\prime} \to W W $) \\
  \hline
  380 & 378 & 4000 & $ 0.071$
%      & xxx & xxx
      & 593 & 144  & $ 97 \%$ \\
  \hline
  500 & 498 & 4000 &$ 0.054$
%      & xxx & xxx
      & 178  & 42 &   $ 99 \%$ \\
  \hline
  \end{tabular}
}
  \caption{\small Representative points in this study. The $\sigma(Z^\prime W)$ 
                  and $\sigma(Z^\prime qq)$ are the $pp \to Z^\prime W $ 
                  production cross section and $pp \to Z^\prime qq$ production 
                  cross section, respectively.}
  \label{tab:representative points}
\end{table*}
%===============================================================================
For the Monte Carlo simulation, we have produced the signal parton events by 
Madgraph/Madevent \cite{Alwall:2007st, Christensen:2009jx} and they have been hadronized by 
PYTHIA \cite{Sjostrand:2003wg}. For the SM backgrounds, we have generated
the events using ALPGEN \cite{Mangano:2002ea} and HERWIG 
\cite{Corcella:2000bw, Corcella:2002jc}. 
Our detector simulation is based on ACERDET \cite{RichterWas:2002ch}.
Hereafter we assume an integrated luminosity of ${\cal L}=100$ fb$^{-1}$.

\begin{figure}[t]
  \begin{center}
    \includegraphics[origin=b, angle=0,width=6cm]{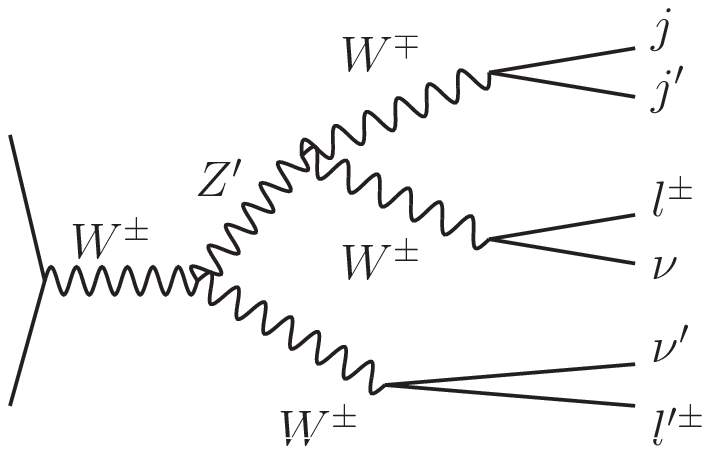}
    \includegraphics[origin=b, angle=0,width=6cm]{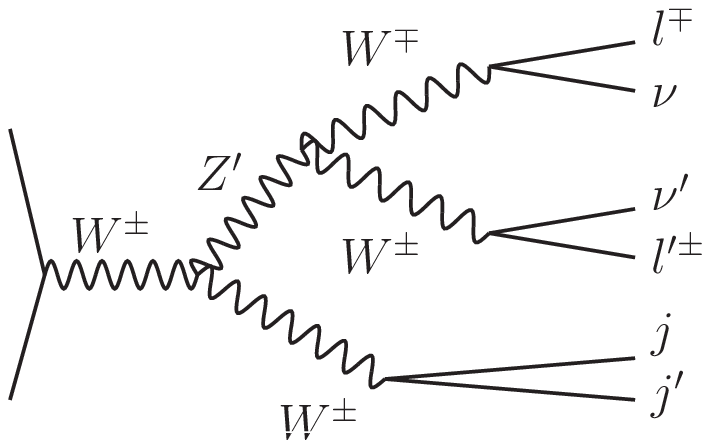}
    \caption{(Left) Feynman diagram for the same sign dilepton mode.
(Right) Feynman diagram for the opposite sign dilepton mode.
} 
   \label{fig:wwz}
  \end{center}
\end{figure}

In our analysis, we concentrate on the $W$-associated production depicted 
in Fig. \ref{fig:wwz}, especially in the following dilepton mode,
\begin{eqnarray}
pp\to W^\pm Z'  \to  \left\{
\begin{array}{ll}
W^\pm\  (W^+ W^-) \to (l^\pm \nu) ((l^\pm\nu) (jj))~~(\mathrm{same~sign}),\\
W^\pm\  (W^+ W^-) \to (jj) ((l^\pm \nu) (l^\mp\nu))~~(\mathrm{opposite~sign}), 
\end{array}
\right.
\end{eqnarray}
because the large SM backgrounds can be reduced in the dilepton modes. 
In particular, it is shown that the $t\bar{t}$ background is significantly 
reduced for the same sign dilepton mode \cite{Han:2009qr}.  

%--------------------------------------------------------->>>FIG

%===================================================================================================
%===================================================================================================
%===================================================================================================

In the signal mode, there are three $W$ bosons; 
two $W$ bosons decay leptonically and one $W$ decays hadronically. 
In order to reconstruct the hadronically decaying $W$,
we require the following cuts:
\begin{itemize}
\item at least two and less than three hard jets with $p_T>20$~GeV,
\item 65~GeV $< m_{jj} < $ 95~GeV,
\end{itemize}
where the $m_{jj}$ is the invariant mass of two hard jets.
We also impose $b$-jet veto to reduce $t \bar{t}$ events.

\begin{figure}[t]
  \begin{center}
    \includegraphics[origin=b, angle=0,width=8cm]{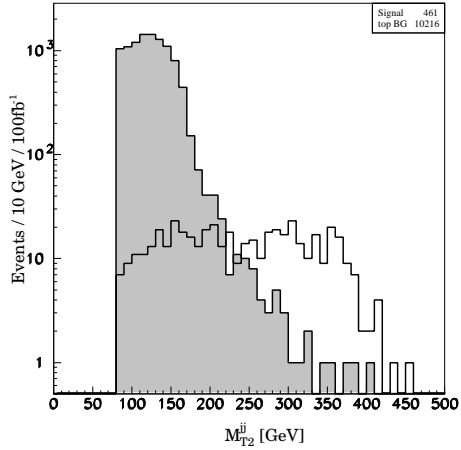}
    \caption{The $M_{T2}^{jj}$ distribution in the opposite sign dilepton mode
for $m_{Z^{\prime}} = 380$ ~GeV (white) and the $t\overline{t}$
 backgrounds (hatched). Here the integrated luminosity is $100$ fb$^{-1}$
and the mass of the missing particle is set to $m_W$.} 
   \label{fig:Mt2jj}
  \end{center}
\end{figure}

To reconstruct the two $W$ bosons which decay leptonically, 
we use neutrino MAOS momenta which are defined by a $M_{T2}$ 
valuable using two leptons and missing momenta. 
In the same (opposite) sign charge mode, we require the following:
\begin{itemize}
%\item less than three leptons ($p_T > 10$~GeV),
\item two leptons with the same (opposite) charge with $p_T > 10$~GeV,
\item  $\ptsl > 50$~GeV.
\end{itemize}
In the same sign dilepton mode, the requirement of the same charge lepton 
reduce the $t \bar{t}$ events significantly.
On the other hand, huge $t \bar{t}$ events still remain for the opposite
sign dilepton mode. To reduce the $t \bar{t}$ events, 
we also define another $M_{T2}$ valuable, $M_{T2}^{jj}$, in which 
we define the "effective missing $p_T$", 
${\psll}^{\rm eff}_T = \psl_T + {\bf p}_{T l1} + {\bf p}_{Tl2}$ and 
the mass of missing particle, $m_X = m_W$, in the opposite charge mode.
The $M_{T2}^{jj}$ valuable should have a maximum value at the top quark mass 
taking the missing particle mass, $m_X = m_W$, because we deal with 
two $W$ bosons decaying leptonically
as missing particles. 
In Fig. \ref{fig:Mt2jj}, we show the $M_{T2}^{jj}(m_X = m_W)$
distributions for the signal with $m_{Z'}=380$ GeV and the $t\overline{t}$
events after the jet and lepton cuts. 
We can see a sharp endpoint around the top quark mass for
the $t\overline{t}$ events while the signal events give larger 
$M_{T2}^{jj}$.
Therefore, we impose $M_{T2}^{jj}(m_X = m_W) > 200$~GeV
to cut the $t \bar{t}$ events. The procedure reduces the combinatorial 
ambiguity of the jets and the leptons than a procedure using $M_{T2}^{jljl}$ 
which should have a maximum value at the top quark mass taking the missing 
particle mass, $m_X = 0 ~\rm{GeV}$. 
%--------------------------------------------------------->>>FIG
%---------------------------------------------------------<<<FIG

In addition, we also impose other cuts, the value of 
$M_\mathrm{eff}$, which is defined as a scalar sum of the visible and
the missing momenta \cite{Hinchliffe:1996iu},
and invariant mass of two leptons, $m_{ll}$, listed in Table \ref{tab:CUTs}. 
The $m_{ll}$ cuts reduce the $Z$ boson and $t \bar{t}$ backgrounds because
a lepton from a top quark decay is softer than a lepton in the signal events.
In Table \ref{table:cut_same} (\ref{table:cut_odd}), we show the
number of events after the selection cuts for the same (opposite) sign mode.

From the MAOS momenta, we can reconstruct the invariant mass of the $Z'$ 
as follows:
\begin{eqnarray}
m^2_{Z'}=  \left\{
\begin{array}{ll}
(p+k^{MAOS}+p_{j_1}+p_{j_2})^2~~~(\mathrm{same~sign}), \\
(p+k^{MAOS}+p'+k'^{MAOS})^2~~(\mathrm{opposite~sign}).
\end{array}
\right.
\end{eqnarray}
In both modes, there is a two-fold ambiguity to reconstruct the $Z^\prime$ 
invariant mass from two $W$ bosons. 
In the same sign mode, there are two choices to combine the lepton momenta 
with the hadron momenta. We choose the combination which gives the smaller 
$m_{Z'}$ in our analysis.
On the other hand, we reconstruct $Z^\prime$ invariant mass from two $W$ 
bosons which decay leptonically in order to confirm the charge of 
the $Z^\prime$ in the opposite sign mode.

\begin{figure}[t]
  \begin{center}
    \includegraphics[origin=b, angle=0,width=5cm]{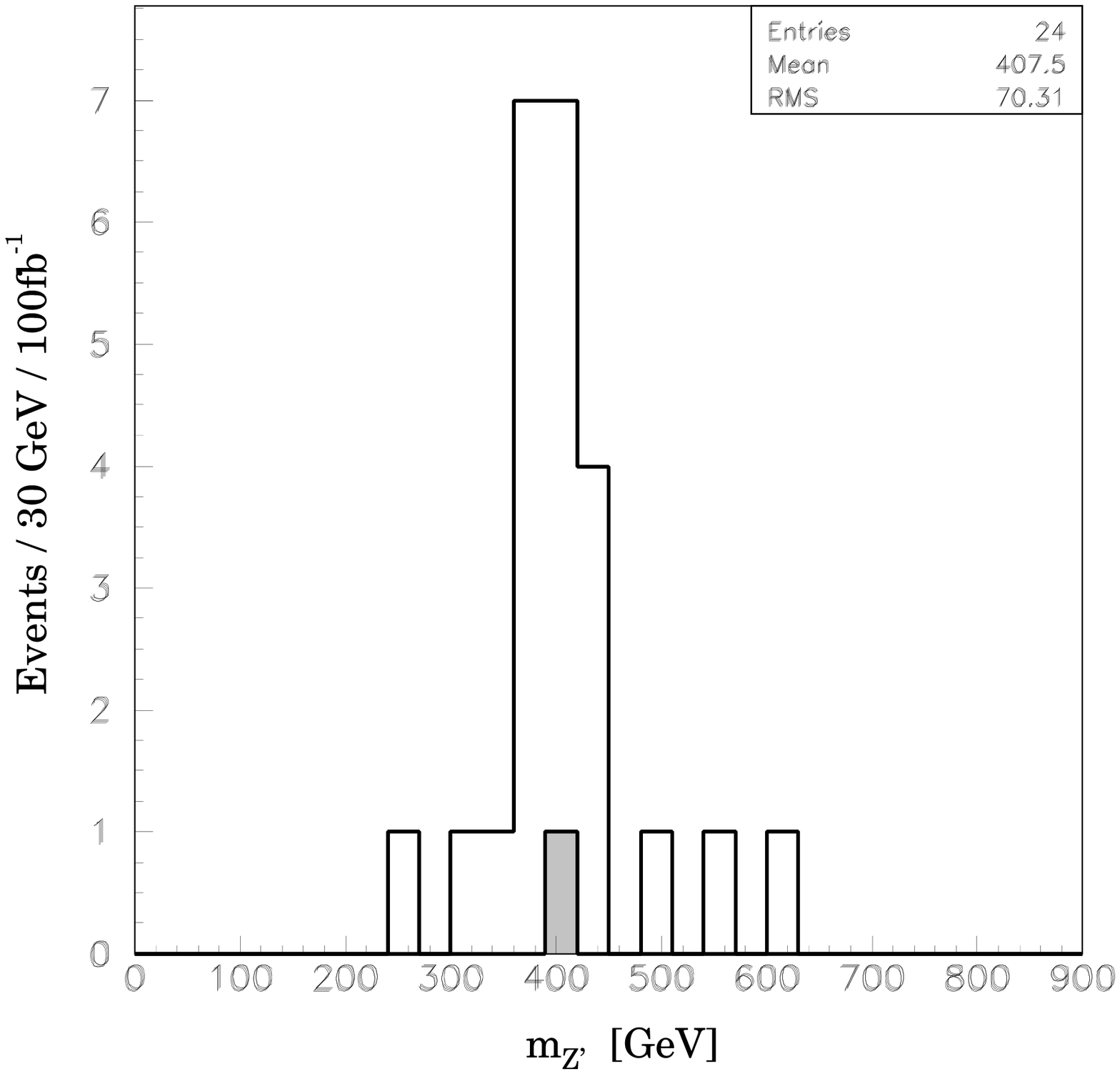}
    \includegraphics[origin=b, angle=0,width=5cm]{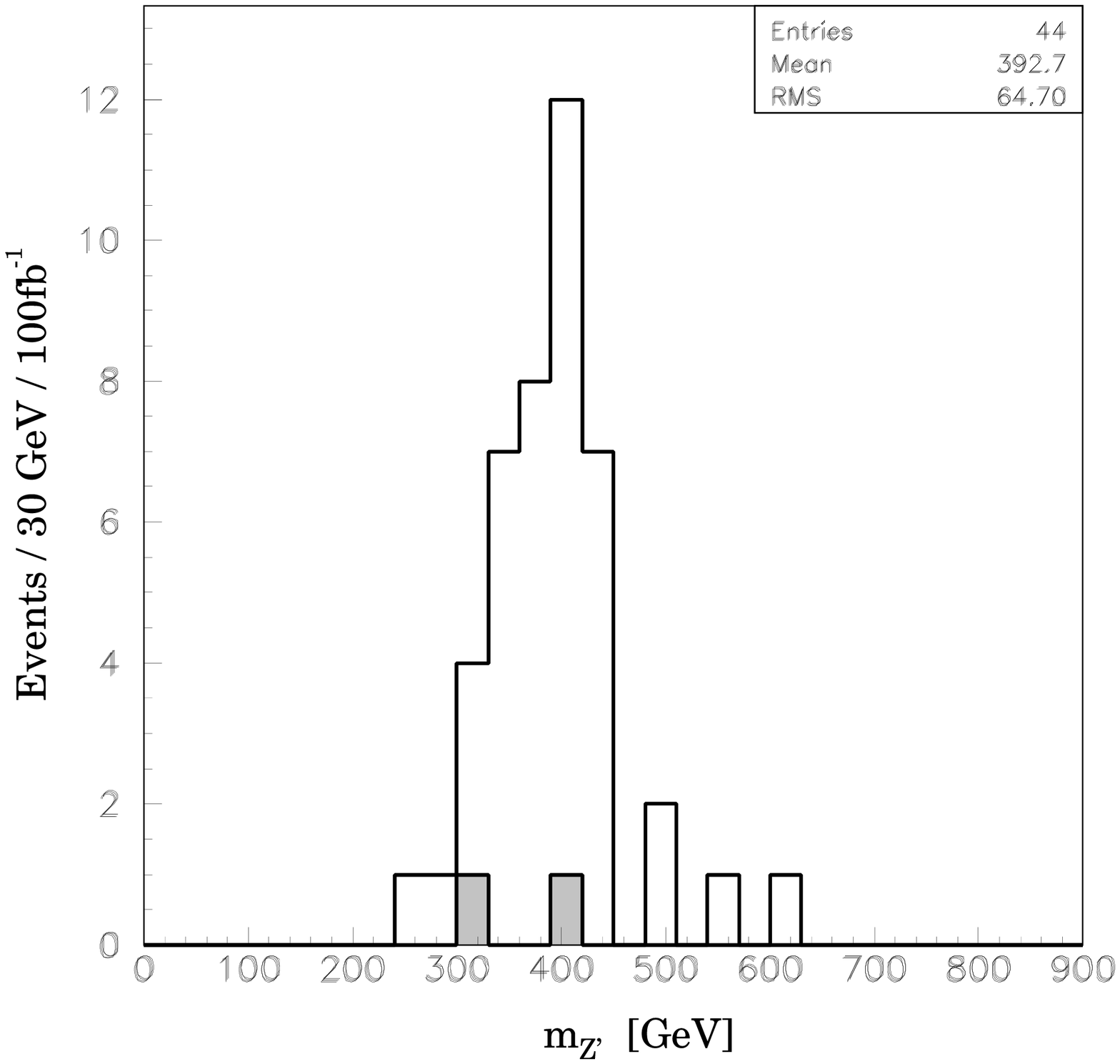}
\\
    \includegraphics[origin=b, angle=0,width=5cm]{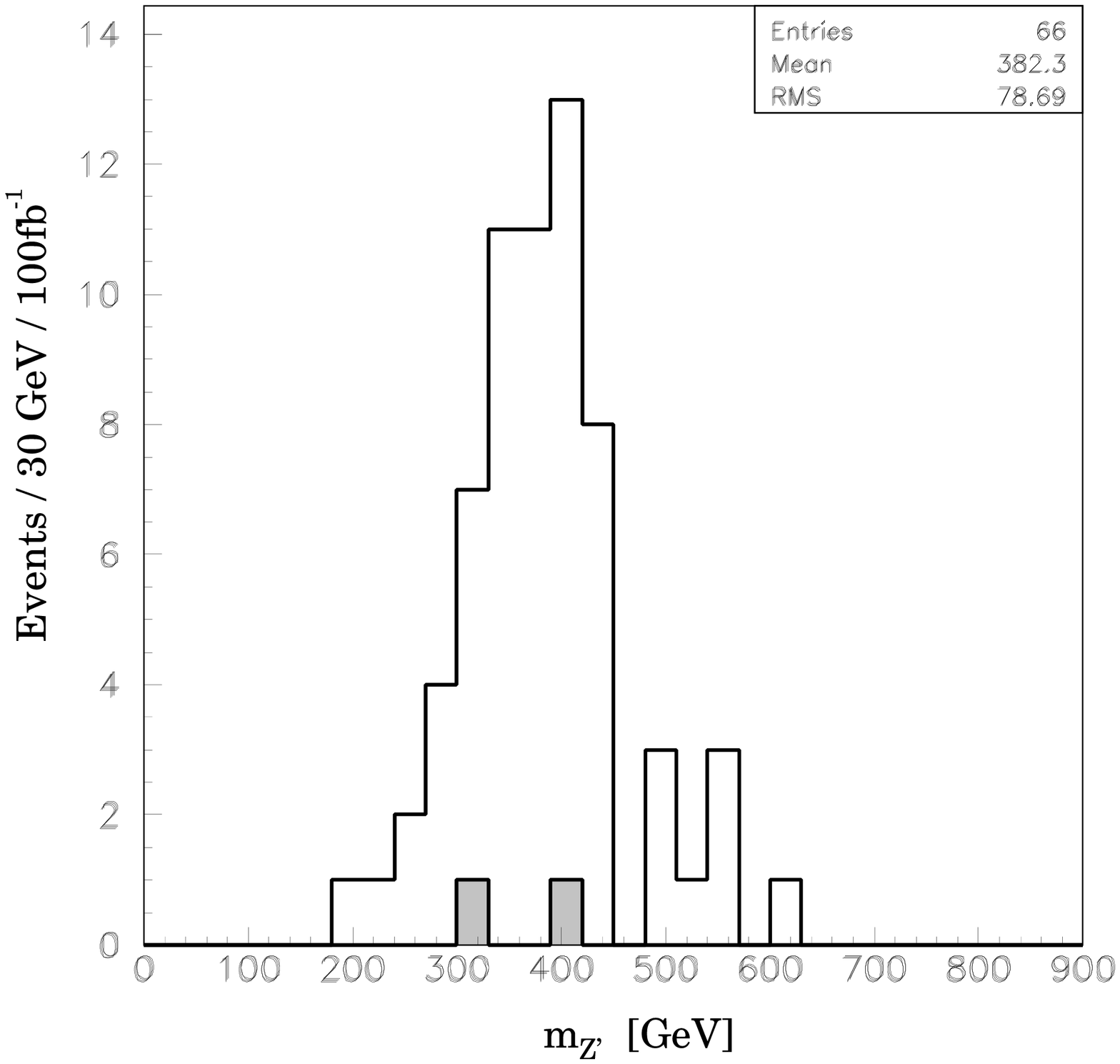}
    \includegraphics[origin=b, angle=0,width=5cm]{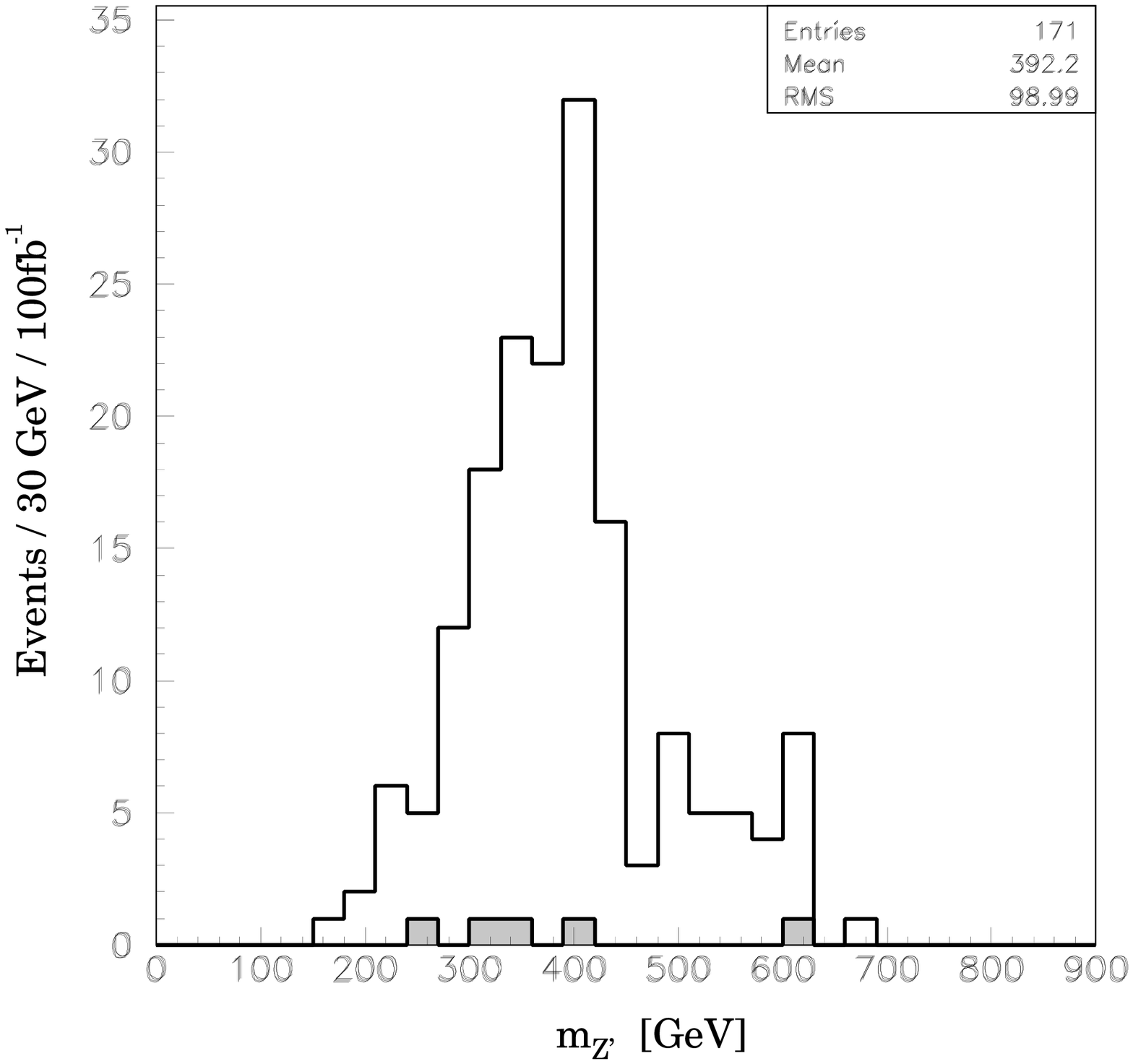}
    \caption{The invariant mass distributions of $Z'$ 
in the same sign dilepton mode for $m_{Z^{\prime}} = 380$~GeV with 
an integrated luminosity ${\cal L}=100^{-1}$ fb. Here we impose the 
cuts listed in Table \ref{table:cut_same} and  
$M_{T2} >$ 60(top left), 40(top right), 20(bottom left), 0(bottom right)~GeV.
The hatched histograms are the SM backgrounds.} 
   \label{fig:mZp380}
  \end{center}
\end{figure}

\begin{figure}[t]
  \begin{center}
    \includegraphics[origin=b, angle=0,width=5cm]{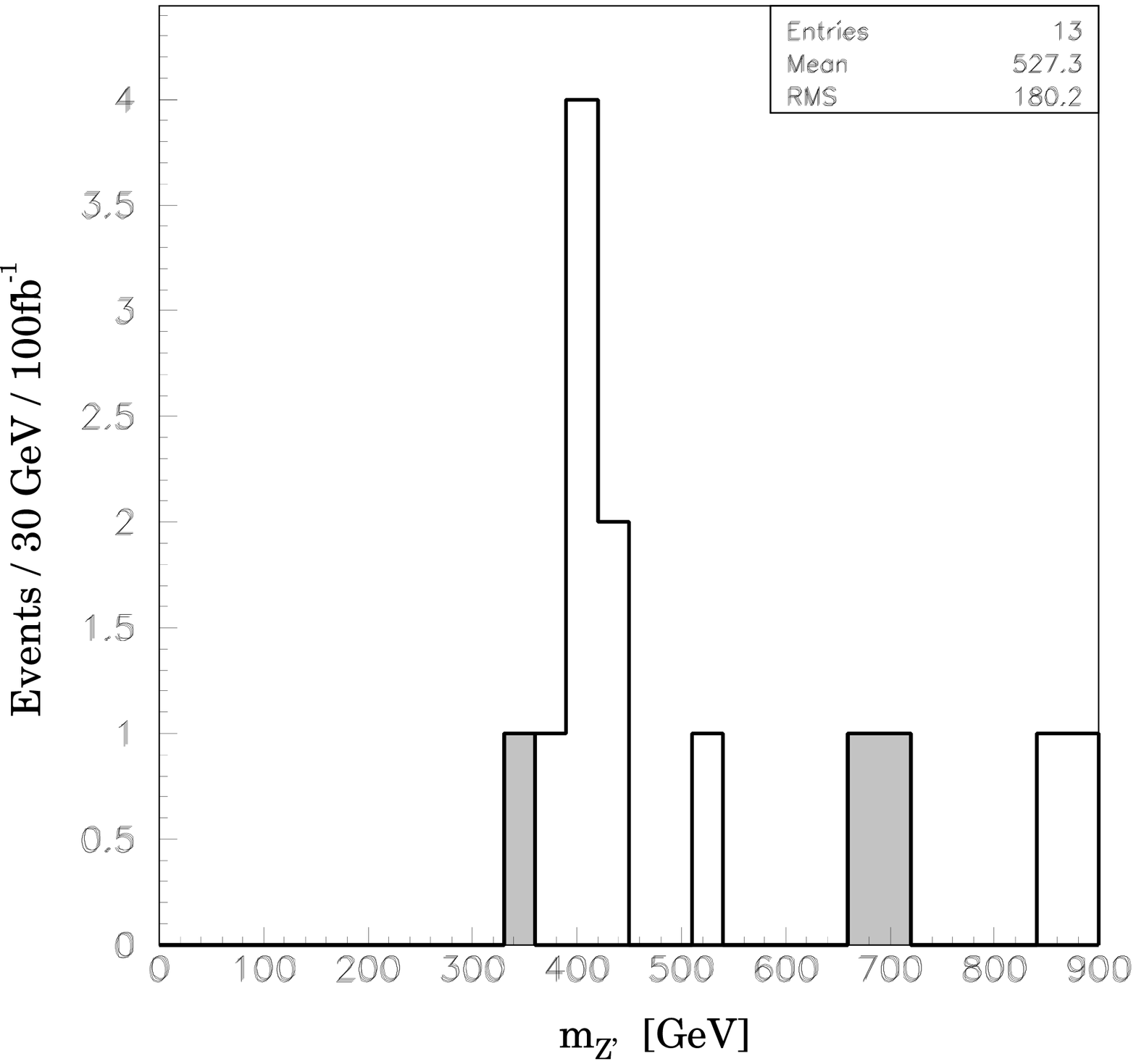}
    \includegraphics[origin=b, angle=0,width=5cm]{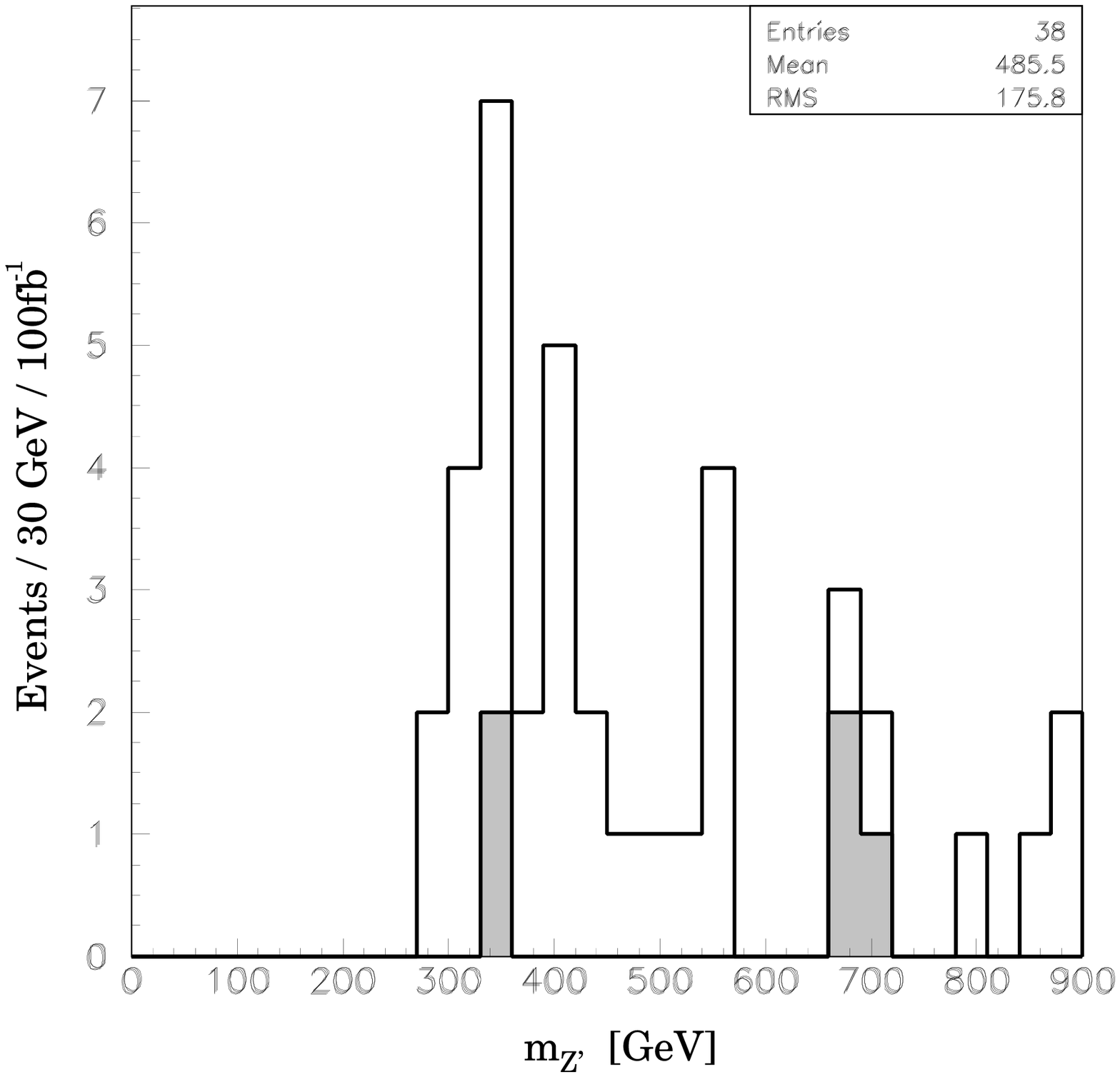}
\\
    \includegraphics[origin=b, angle=0,width=5cm]{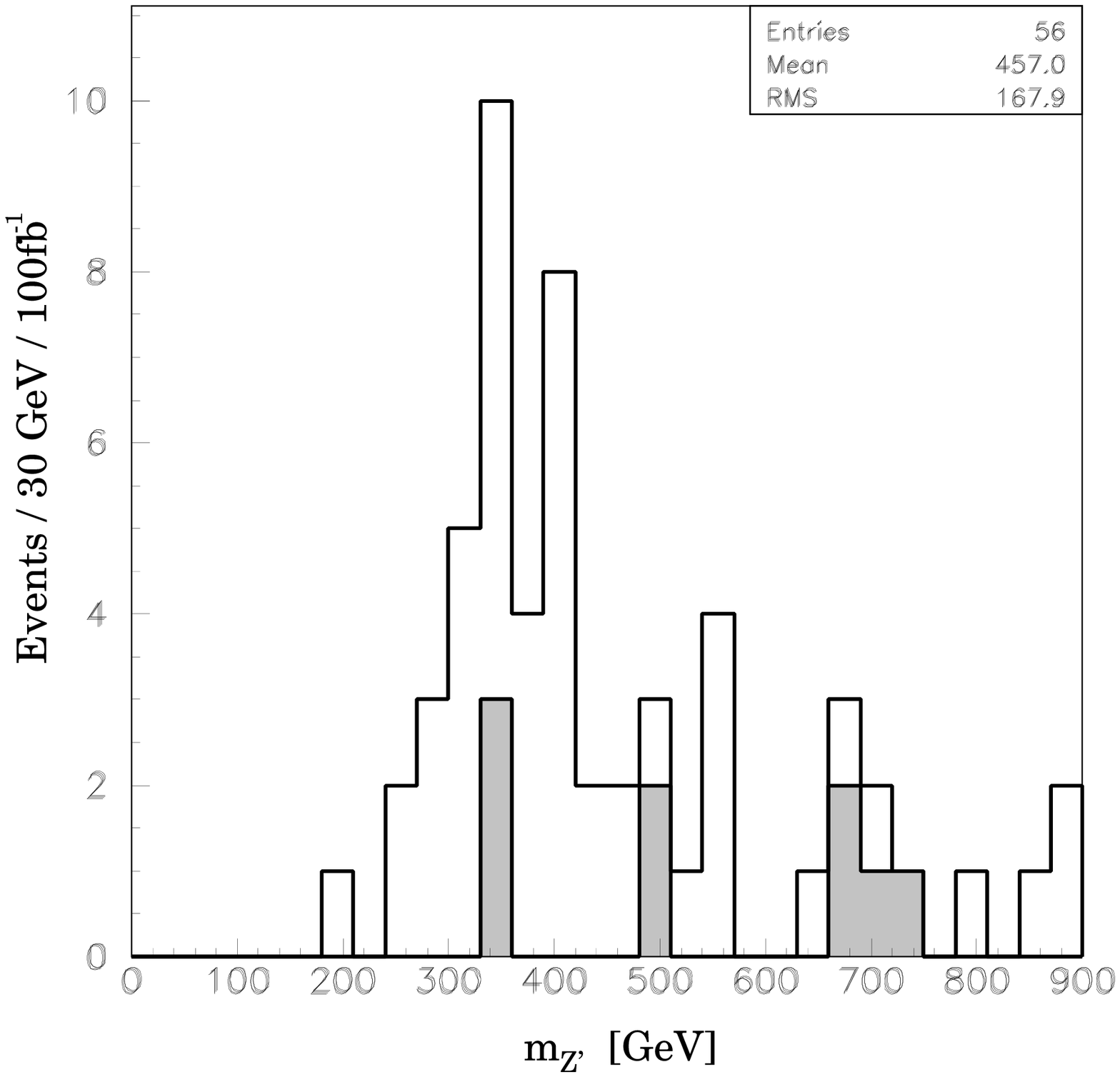}
    \includegraphics[origin=b, angle=0,width=5cm]{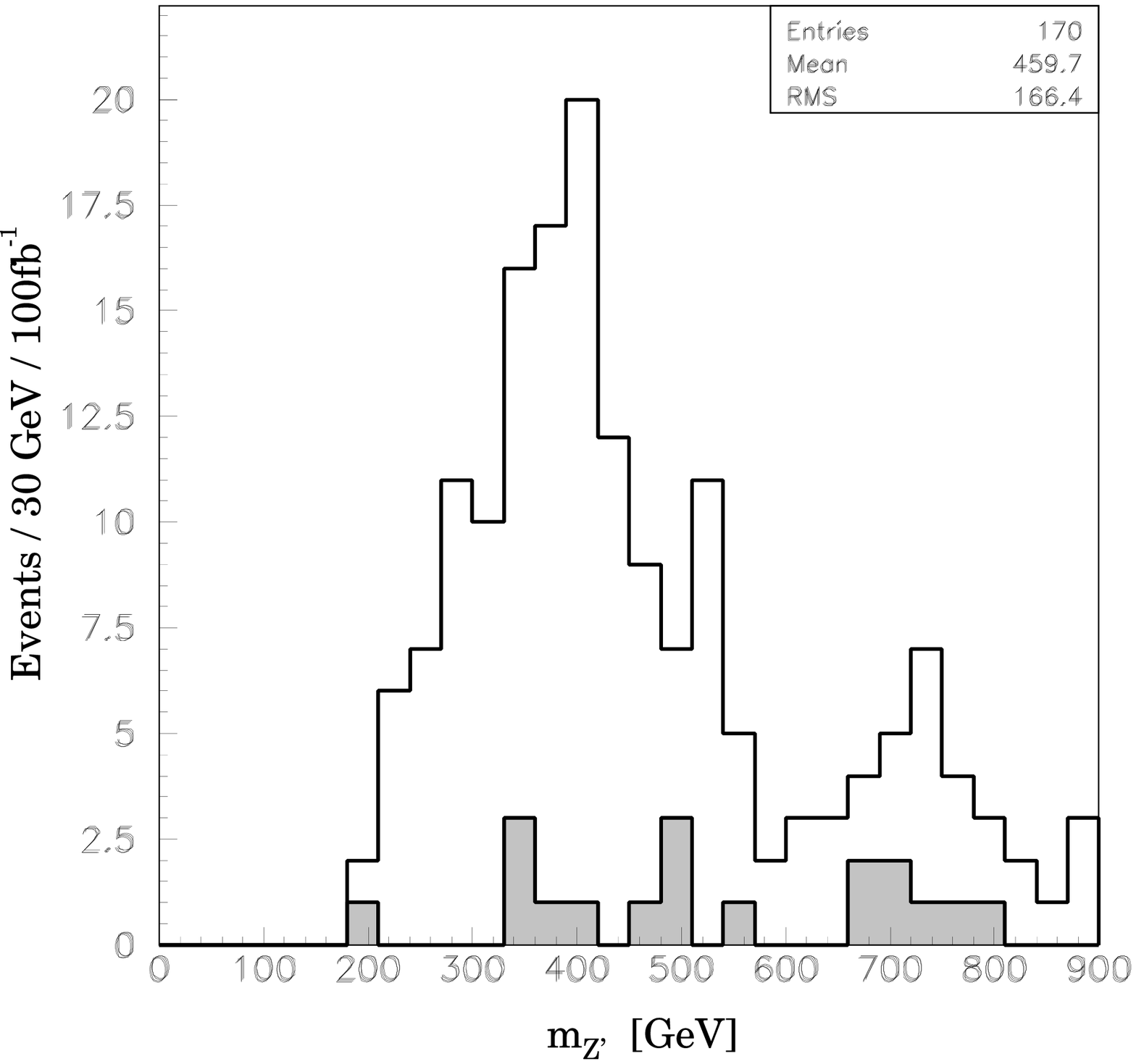}
    \caption{The invariant mass distributions in the opposite sign dilepton mode
for $m_{Z^{\prime}} = 380$~GeV with an integrated luminosity 
${\cal L}=100^{-1}$ fb. Here we impose the 
cuts listed in Table \ref{table:cut_odd} and
$M_{T2} >$ 60(top left), 40(top right), 20(bottom left), 0(bottom right)~GeV.
The hatched histograms are the SM backgrounds.} 
   \label{fig:mZp380_odd}
  \end{center}
\end{figure}

In Figs. \ref{fig:mZp380} and \ref{fig:mZp380_odd},
we show the invariant mass distributions of $Z'$ in the same and opposite 
sign dilepton mode for $m_{Z'}=380$ GeV, respectively. The hatched histograms 
are the SM backgrounds. Although the MAOS momenta become equal to the true 
neutrino momenta only at the $M_{T2}$ endpoint, $M_{T2}\simeq 80$ GeV, we use 
the MAOS momenta in wide $M_{T2}$ regions and plot the invariant mass 
distributions of $Z'$ for $M_{T2}>60, 40, 20, 0$ GeV. 
\footnote{
When the $Z^\prime$ is heavy enough to decay to two on-shell $W$ bosons, 
there is the other definition of the MAOS momenta \cite{Cho:2008tj}.
Even if we take the MAOS momenta, the distributions are not drastically changed
in our study with $100$ fb$^{-1}$.
}
In Fig. \ref{fig:mZp380} we can see the peaks around the true $Z'$ mass 
regions for all the cases. The shape becomes sharper when we use the MAOS 
momenta in the higher $M_{T2}$ regions. However, the number of the events 
becomes smaller. 
Thanks to the $M_{T2}^{jj}$ cut, we can see clear signal peaks around the true 
$m_{Z'}$ value even in the opposite sign mode. Although the shapes of the 
invariant distributions are slightly irregular compared with the same sign mode, 
we can determine the $Z'$ mass from the peak.

\begin{figure}[t]
  \begin{center}
    \includegraphics[origin=b, angle=0,width=5cm]{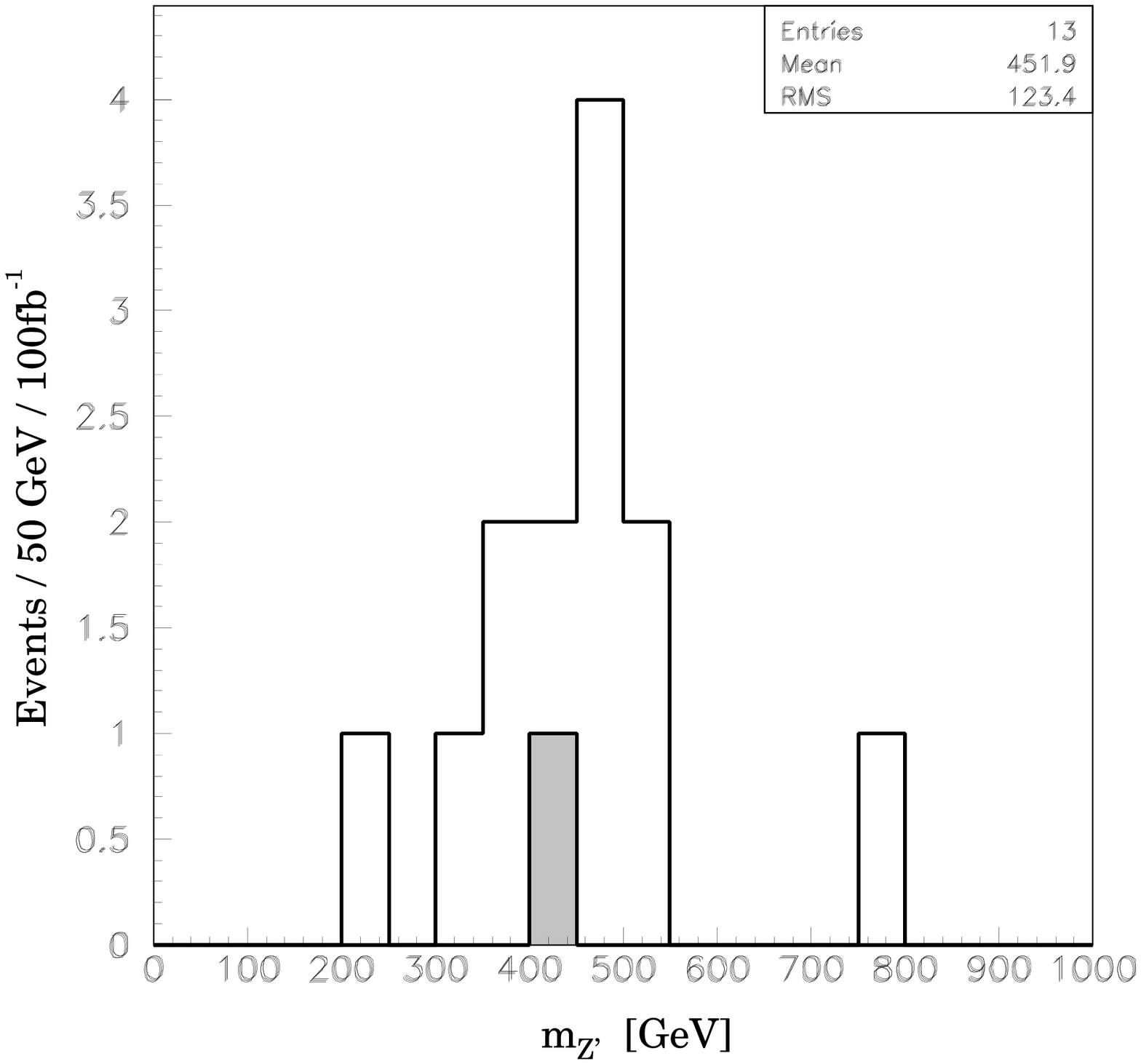}
    \includegraphics[origin=b, angle=0,width=5cm]{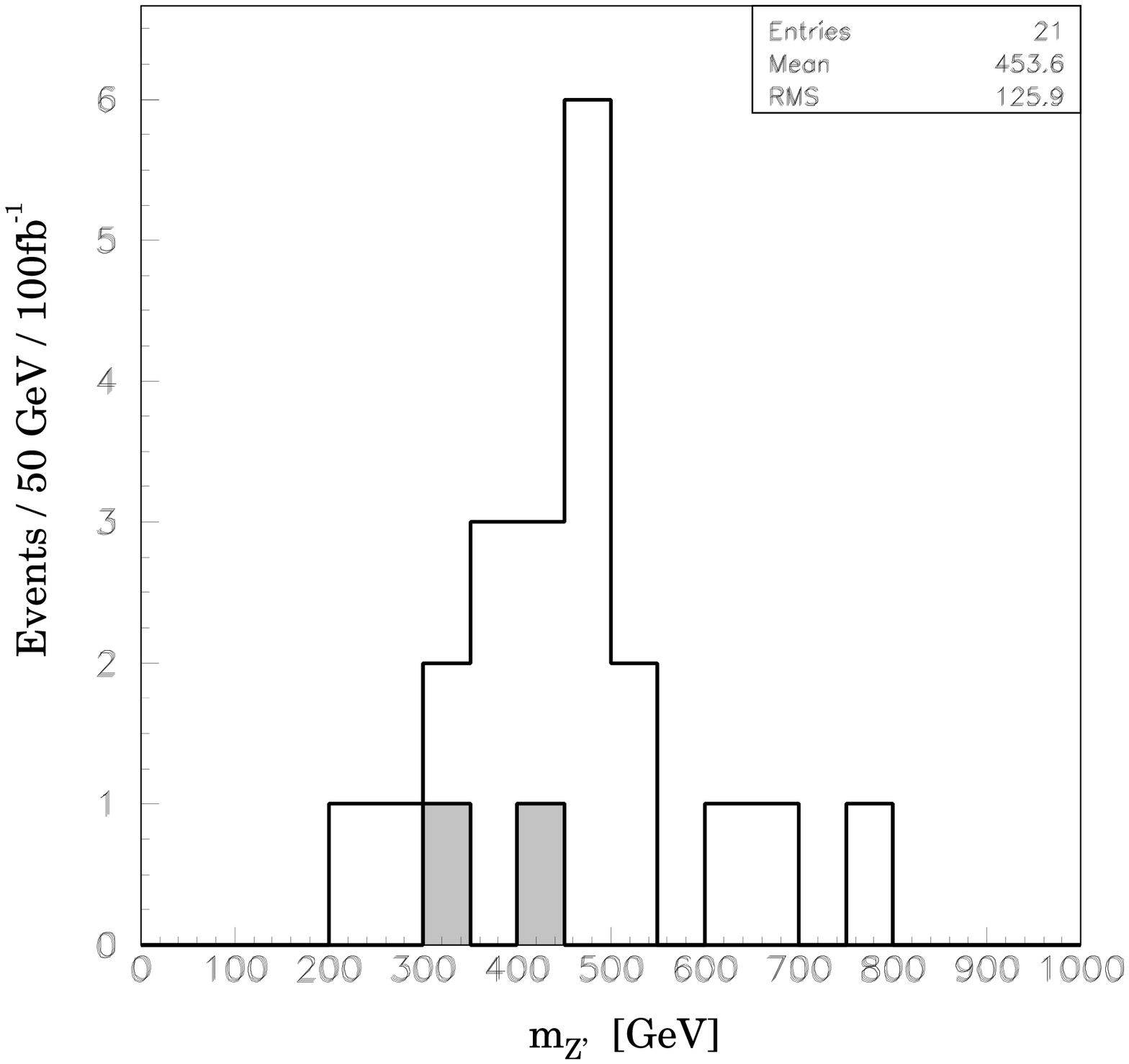}
\\
    \includegraphics[origin=b, angle=0,width=5cm]{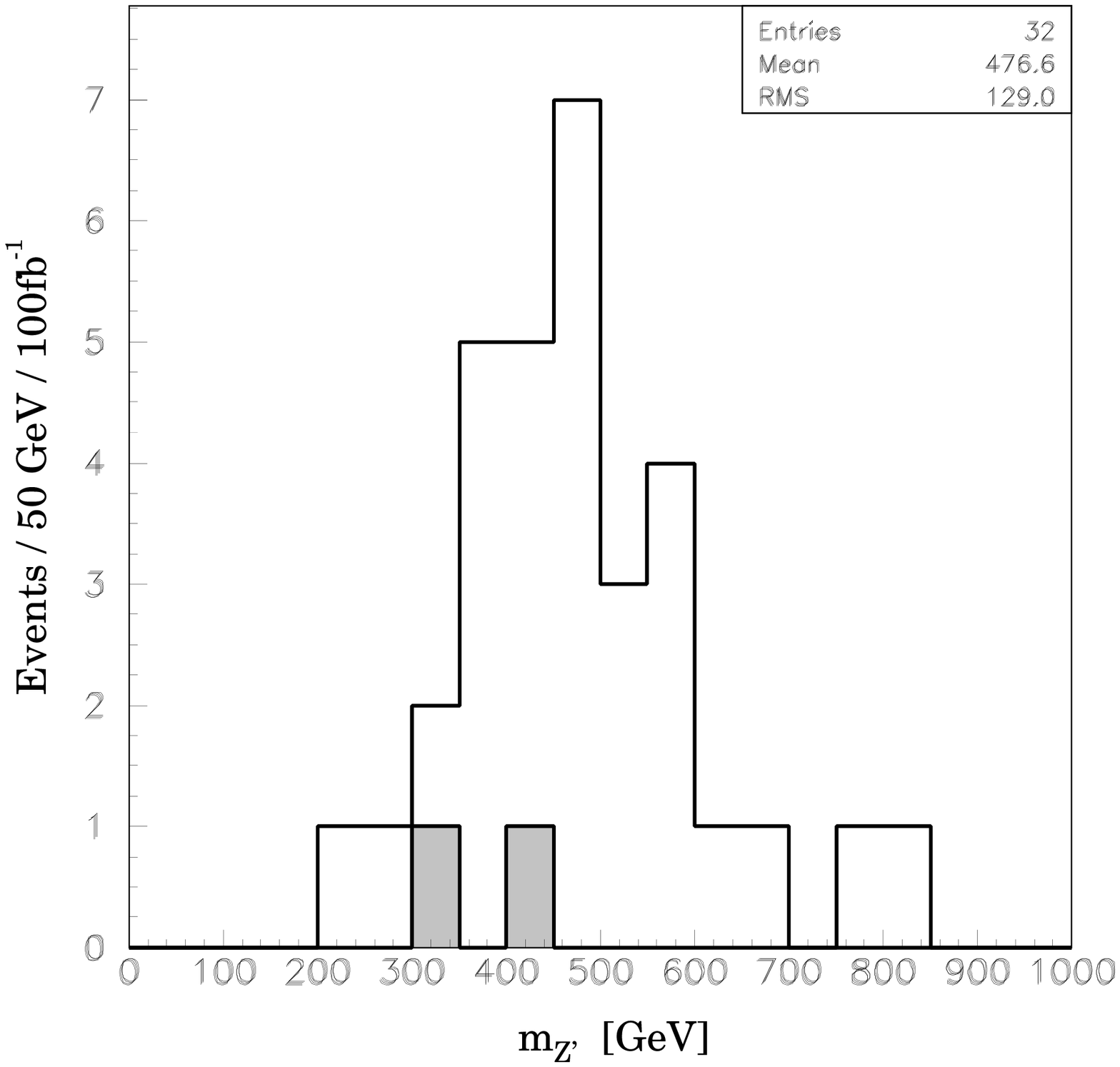}
    \includegraphics[origin=b, angle=0,width=5cm]{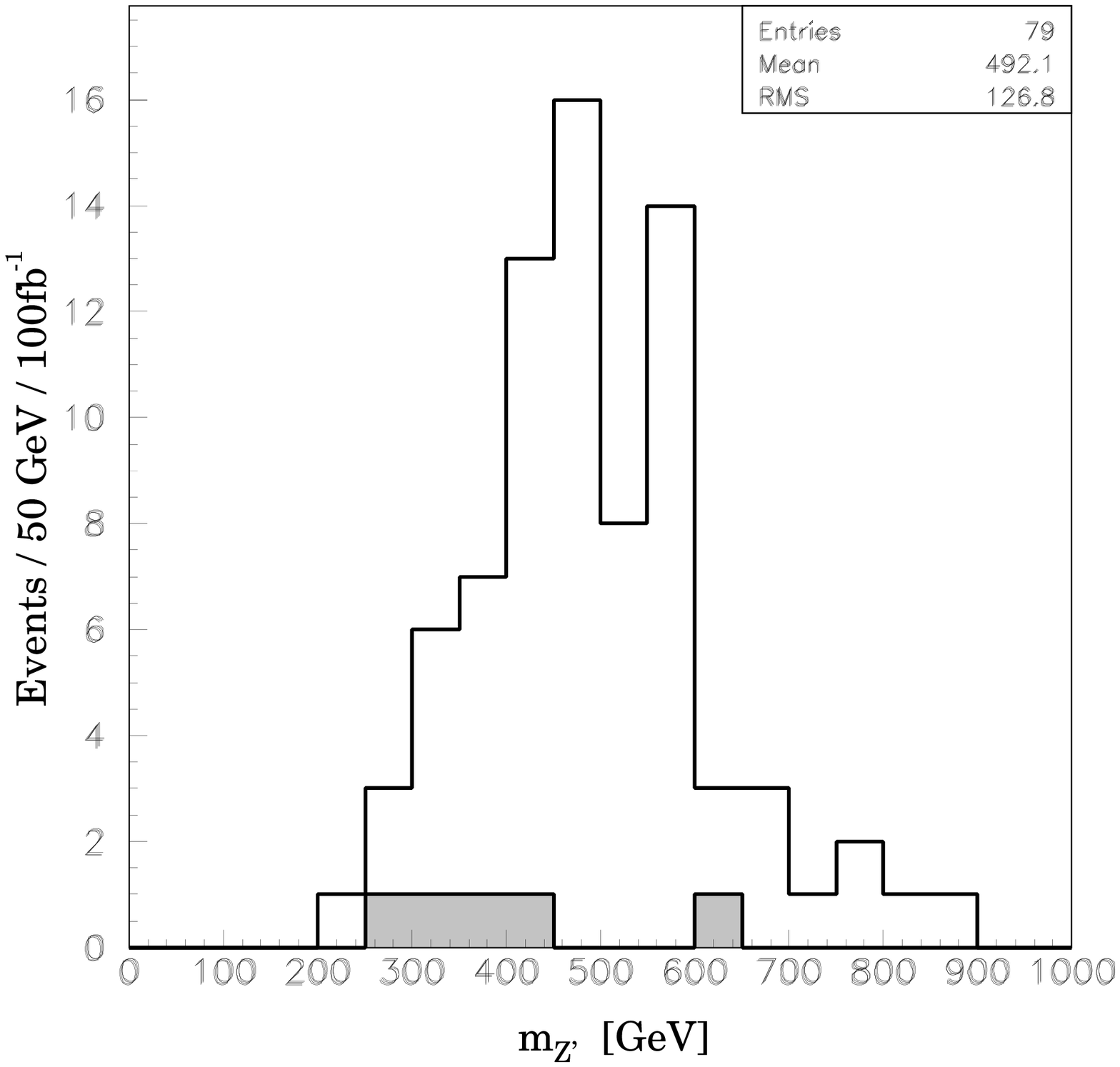}
    \caption{The invariant mass distributions in the same sign dilepton mode
for $m_{Z^{\prime}} = 500$~GeV with an integrated luminosity 
${\cal L}=100^{-1}$ fb.  Here we impose the cuts listed in 
Table \ref{table:cut_same} and 
$M_{T2} >$ 60(top left), 40(top right), 20(bottom left), 0(bottom right)~GeV.
The hatched histograms are the SM backgrounds.} 
   \label{fig:mZp500}
  \end{center}
\end{figure}

\begin{figure}[t]
  \begin{center}
    \includegraphics[origin=b, angle=0,width=5cm]{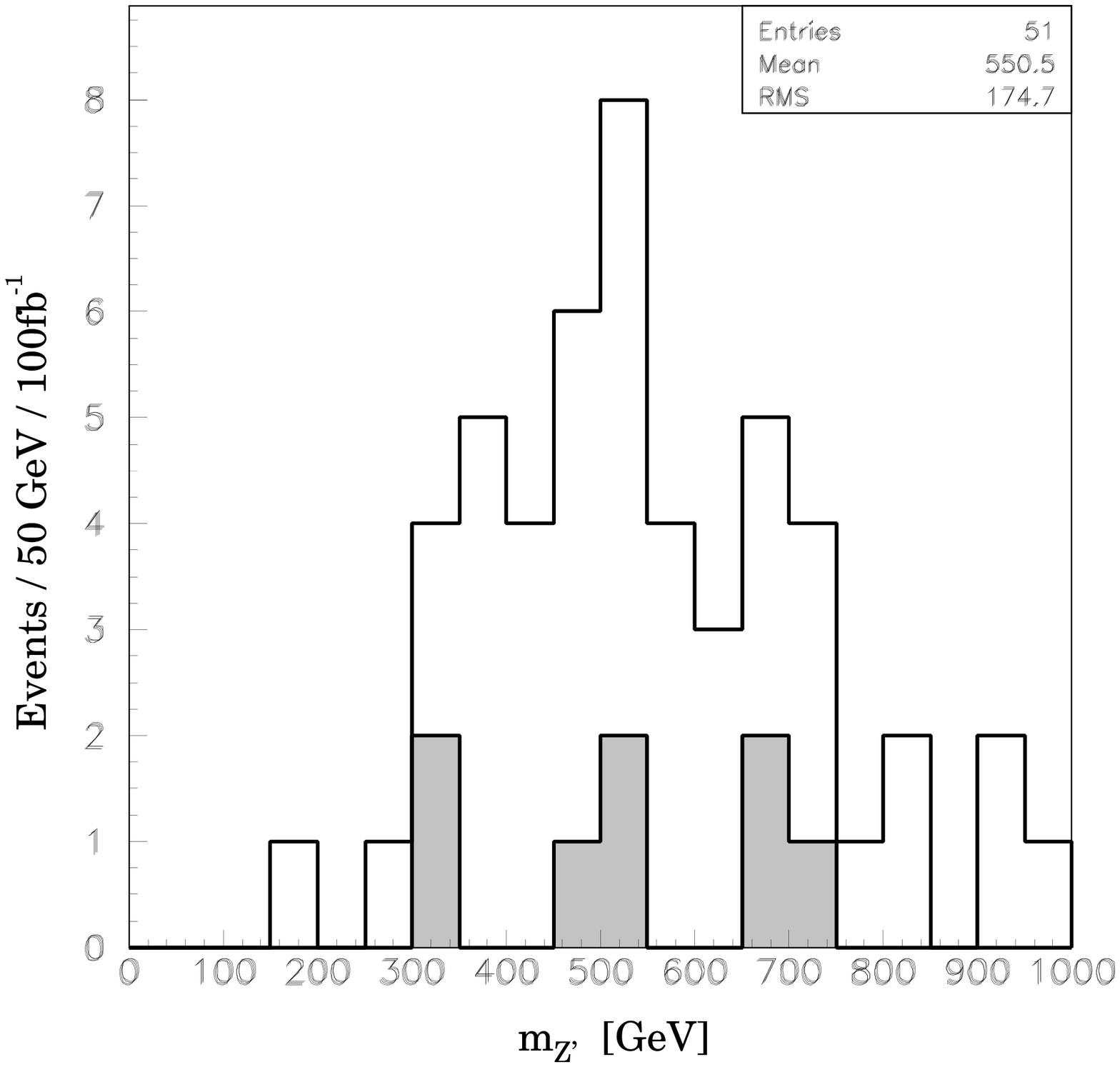}
    \caption{The invariant mass distributions in the opposite sign dilepton mode
for $m_{Z^{\prime}} = 500$~GeV with an integrated luminosity ${\cal L}=100^{-1}$ fb. Here we take the cuts listed in Table \ref{table:cut_odd} and 
$M_{T2} >0$~GeV. The hatched histograms are the SM backgrounds.} 
   \label{fig:mZp500_odd}
  \end{center}
\end{figure}

In Figs. \ref{fig:mZp500} and \ref{fig:mZp500_odd},
we show the invariant mass distributions of $Z'$ in the same and opposite sign 
dilepton mode for $m_{Z'}=500$ GeV, respectively.
Since the production cross section is smaller compared with the $m_{Z'}=380$ GeV 
case, the numbers of events  are limited if we use the tight $M_{T2}$ cut. 
In the opposite sign mode, we cannot see clear peaks for the $M_{T2}>20, 40, 60$ 
GeV cases. However, we can see the peaks around the true $m_{Z'}$ region even for 
$m_{Z'}=500$ GeV in the both modes.

%===============================================================================
\begin{table*}[t]
\center{
  \begin{tabular}{|c||c|c|}
  \hline
                     & $m_{Z^{\prime}} = 380$~GeV & $m_{Z^{\prime}} = 500$~GeV \\
  \hline
   same & $M_\mathrm{eff}>500$~GeV and $m_{ll}>130$~GeV & $M_\mathrm{eff}>500$~GeV and $m_{ll}>130$~GeV \\
  \hline
  opposite & $M_\mathrm{eff}>600$~GeV and $m_{ll}>110$~GeV & $M_\mathrm{eff}>700$~GeV and $m_{ll}>110$~GeV \\
  \hline
  \end{tabular}
}
  \caption{\small Cuts used in the analysis for the same and the opposite sign lepton modes.}
  \label{tab:CUTs}
\end{table*}
%===============================================================================

\begin{table}
\begin{center}
                \begin{tabular}{|c|l|c|r|r|r|} \hline
          & Selection cut  & Signal                & top BG & W boson BG \\
          &                & (380, 500 GeV)        &        &            \\ \hline\hline
jet        & \# of jets ($P_T > 20$~GeV)=2   & (4844, 1451)    & 774624   & 5862    \\ \cline{2-5}
           & 65~GeV $< m_{jj} < $ 95~GeV     & (1116, 313)    & 124142   & 1485    \\ \cline{2-5}
           & b veto                          & (1069, 299)    &  44009   & 1410    \\ \hline\hline

lepton     & 2 same charge leptons ($P_T > 10$~GeV)&  (284, 96)    &   1234   &   91    \\ \cline{2-5}
           & $\ptsl> 50$~GeV                 &  (257, 90)    &    777   &   64    \\ \hline\hline

other      & $M_\mathrm{eff}>500$~GeV        &  (207, 85) &  21   &  9    \\ \cline{2-5}
           & $m_{ll}>130$~GeV                &  (167, 74) &   0   &  5    \\ \hline
                \end{tabular}
\caption{Number of events after selection cuts for the same sign mode with an integrated luminosity ${\cal L}=100^{-1}$ fb.}
\label{table:cut_same}
\end{center}
\end{table}

\begin{table}
\begin{center}
                \begin{tabular}{|c|l|c|r|r|r|} \hline
          & Selection cut  & Signal                & top BG & W boson BG \\
          &                & (380, 500 GeV)        &        &            \\ \hline\hline
jet        & \# of jets ($P_T > 20$~GeV)=2   & (4844, 1451)    & 774624   & 5862    \\ \cline{2-5}
           & 65~GeV $< m_{jj} < $ 95~GeV     & (1116, 313)    & 124142   & 1485    \\ \cline{2-5}
           & b veto                          & (1069, 299)    &  44009   & 1410    \\ \hline\hline

lepton     & 2 opposite charge leptons ($P_T > 10$~GeV)&  (508, 141)    &  15845   &  993    \\ \cline{2-5}
           & $\ptsl> 50$~GeV                 &  (461, 130)    &  10216   &  533    \\ \hline\hline

$M_{T2}^{jj}$   & $M_{T2}^{jj}$                        &  (289, 99)    &    125   &   44    \\ \hline\hline

other      & $M_\mathrm{eff}> (600, 700)$~GeV&  (254, 83) &  (25, 9)   &  (16, 12)    \\ \cline{2-5}
           & $m_{ll}>110$~GeV                &  (206, 67) &   (9, 1)   &  (11, 8)    \\ \hline
                \end{tabular}
\caption{Number of events after selection cuts for the opposite 
sign mode with an integrated luminosity ${\cal L}=100^{-1}$ fb.}
\label{table:cut_odd}
\end{center}
\end{table}

%--------------------------------------------------------->>>FIG
%---------------------------------------------------------<<<FIG

%--------------------------------------------------------->>>FIG
%---------------------------------------------------------<<<FIG

%--------------------------------------------------------->>>FIG
%---------------------------------------------------------<<<FIG

%--------------------------------------------------------->>>FIG
%---------------------------------------------------------<<<FIG

%%%%%%%%%%%%%%%%%%%%%%%%%%%%%%%%%%%%%%%%%%%%%%%%%%%%%%%%%%%%%%%%%%%%%%%%%%%%%%%%%
\section{ Summary and Discussion}\label{sect:Summary}
%%%%%%%%%%%%%%%%%%%%%%%%%%%%%%%%%%%%%%%%%%%%%%%%%%%%%%%%%%%%%%%%%%%%%%%%%%%%%%%%%
In this paper, we have studied the search for the neutral vector resonance,
focusing on a fermiophobic case at the LHC. Such fermiophobic vector bosons 
are predicted by perturbative TeV scale scenarios without the Higgs boson,
and can be produced in the  $W$-associated processes at the LHC.
We have studied the dilepton decay modes,
$pp \to W^{\pm}Z^{\prime} \to W^{\pm}W^{\pm}W^{\mp} 
             \to l \nu l^{\prime} \nu^{\prime} jj$, 
since the SM model backgrounds are suppressed. 
There are two neutrinos in the dilepton modes and we cannot measure
each neutrino momentum directly. 
In order to determine the neutrino momenta we use the MAOS momenta.
From the MAOS momenta we can reconstruct the invariant mass of 
the neutral vector boson and the mass of the neutral vector resonance 
can be determined by the peak of the invariant mass distribution.

We have applied the method to both the same and the opposite sign dilepton 
modes. We have found that the $M_{T2}^{jj}$ cut is very effective to reduce 
the $t\bar{t}$ background, especially in the opposite sign mode.
We can obtain clear invariant mass distributions of the vector resonance
and the shape of the invariant mass distributions changes depending on 
the $M_{T2}$ cut. Although the shape becomes broader if we use a looser $M_{T2}$ 
cut, we have found that the peak positions of the invariant mass distributions 
do not change drastically. To measure the mass more accurately, we should 
optimize the $M_{T2}$ cut analysing detailed Monte Carlo simulations.

Let us discuss advantages of our method compared to other analyses on the $Z'$
measurements at the LHC.  
Since we have studied the dilepton mode in the $Z'$ decay, the SM backgrounds 
are reduced and we can see the clear peak of the $Z'$ resonance.   
From the charge of the dileptons, we can confirm the charge of the resonance
in the opposite sign mode. In the same sign mode, we can also infer that the 
charge is neutral unless double charged particles exist.
If we consider one lepton mode, $Z^\prime W \to l \nu jjjj$ or 
$Z^\prime \to W^+W^- \to l \nu jj$, which is produced via the vector boson fusion or 
Drell-Yan production process, the $Z^{\prime}$ and $W^{\prime}$ resonances may 
overlap each other due to difficulty of the $W$($Z$) identification at the LHC.
However, it is possible to study the $Z'$ peak separately in our method.
In the three site Higgsless model, the $Z^\prime$ can also be measured 
via Drell-Yan production process in the semileptonic decay mode because the 
couplings with SM fermions are not zero even if we take the ideal delocalization
\cite{Ohl:2008ri,Alves:2009aa,Bach:2009zz}. Using MAOS momenta, the 
$Z^\prime$ may also be measured in the leptonic decay mode, 
$Z^\prime \to WW \to l \nu l^\prime \nu^\prime $, without $W^\prime$ 
contamination. The similar signal, 
${\rm Higgs} \to WW \to l \nu l^\prime \nu^\prime $, 
has been studied by using MAOS momenta \cite{Choi:2009hn,Choi:2010dw}.
Combining the results from the $W$-associated and Drell-Yan production process 
study with MAOS momenta, we may also determine the $Z^\prime$ couplings with weak 
gauge bosons and SM fermions.

Although we focus on the $Z^{\prime}$ measurement in the three-site Higgsless model
in this paper, the method will be also useful for the mass measurement
of general fermiophobic neutral vector resonances.

%%%%%%%%%%%%%%%%%%%%%%%%%%%%%%%%%%%%%%%%%%%%%%%%%%%%%%%%%%%%%%%%%%%%%%%%%%%%%%%%%
\section*{ Acknowledgments }
%%%%%%%%%%%%%%%%%%%%%%%%%%%%%%%%%%%%%%%%%%%%%%%%%%%%%%%%%%%%%%%%%%%%%%%%%%%%%%%%%
We would like to thank Masafumi Kurachi for discussions and comments.
This work is supported in part by the Grant-in-Aid for the Global COE Program Weaving Science Web beyond Particle-matter Hierarchy from the Ministry of Education, Culture, Sports, Science and Technology of Japan (M.~A.).

%%%%%%%%%%%%%%%%%%%%%%%%%%%%%%%%%%%%%%%%%%%%%%%%%%%%%%%%%%%%%%%%%%%%%%%%%%%%%%%%%

\end{document}